\documentclass[]{spie}  

\usepackage{amsmath,amsfonts,amssymb}
\usepackage{graphicx}
\usepackage{multirow}
\usepackage{adjustbox}
\usepackage[colorlinks=true, allcolors=blue]{hyperref}

\title{Lesion Focused Super-Resolution}

\author[a]{Jin Zhu}
\author[b,c]{Guang Yang}
\author[a]{Pietro Lio}
\affil[a]{ The Computer Laboratory, University of Cambridge, Cambridge, CB3 0FD, UK}
\affil[b]{ Cardiovascular Research Centre, Royal Brompton Hospital, London, SW3 6NP, UK}
\affil[c]{ National Heart and Lung Institute, Imperial College London, London, SW7 2AZ, UK}

\authorinfo{Further author information: (Send correspondence to Jin Zhu: E-mail: jin.zhu@cl.cam.ac.uk, Guang Yang: E-mail: g.yang@imperial.ac.uk and Pietro Lio: E-mail: pietro.lio@cl.cam.ac.uk)}

\begin{document} 
\maketitle

\begin{abstract}
Super-resolution (SR) for image enhancement has great importance in medical image applications. Broadly speaking, there are two types of SR, one requires multiple low resolution (LR) images from different views of the same object to be reconstructed to the high resolution (HR) output, and the other one relies on the learning from a large amount of training datasets, i.e., LR-HR pairs. In real clinical environment, acquiring images from multi-views is expensive and sometimes infeasible. In this paper, we present a novel Generative Adversarial Networks (GAN) based learning framework to achieve SR from its LR version. By performing simulation based studies on the Multimodal Brain Tumor Segmentation Challenge (BraTS) datasets, we demonstrate the efficacy of our method in application of brain tumor MRI enhancement. Compared to bilinear interpolation and other state-of-the-art SR methods, our model is lesion focused, which is not only resulted in better perceptual image quality without blurring, but also more efficient and directly benefit for the following clinical tasks, e.g., lesion detection and abnormality enhancement. Therefore, we can envisage the application of our SR method to boost image spatial resolution while maintaining crucial diagnostic information for further clinical tasks.
\end{abstract}

\keywords{Super-resolution, lesion detection, medical image analysis, image processing}

\section{INTRODUCTION}
\label{sec:intro}

Images with high resolution (HR) are greatly in demand for many real applications\cite{trinh2014novel}. However, the resolution and quality of the images are normally limited by the imaging hardware. For medical images, which provide useful and crucial details of the anatomical and physiological information for the patients, are very desirable with HR. In addition to the possible restrictions of the imaging hardware, medical images are more susceptible by the health limitations (e.g., ionizing radiation dose of using X-ray) and acquisition time limitations (e.g., Specific Absorption Rate limits of using MRI). Moreover, movements due to patients fatigue and organs pulsation will further degrade image qualities and result in images with lower signal-to-noise ratio (SNR). Low resolution (LR) medical images with limited field of view and degraded image quality could reduce the visibility of vital pathological details and compromise the diagnostic accuracy and prognosis \cite{yang2016combined}.

Research studies have shown that image super-resolution (SR) provides an alternative and relatively cheaper solution to improve the perceptual quality of medical images in terms of the spatial resolution enhancement instead of hardware improvement. Compared to conventional image interpolation, SR methods can provide better HR outputs with higher SNR and less blurry effects. Broadly speaking, there are two different types of SR: (1) using multiple LR images acquired from different views of the same object to reconstruct the HR output, but acquiring multi-view images could be expensive and sometimes infeasible; (2) learning a particular SR model using LR-HR training pairs, and performing the inference on a new input LR image to yield the HR output \cite{yang2012coupled,trinh2014novel}.

More recently, deep learning base SR methods have boosted the performance of the super-resolved HR images mainly owe to the development of the computing power and the available big data. For example, the SRGAN method \cite{ledig2017photo}, which was developed based on a Generative Adversarial Network (GAN) model, has demonstrated fast and accurate SR results. However, SRGAN has been developed for natural images and there are still limited studies for medical images.
%
%

In this study, we developed a lesion focused SR (LFSR) method that leverage the merits of GAN based models to generate perceptually more realistic SR results and also avoid introducing non-existing features into the lesion area after SR. By performing simulation based studies on the Multimodal Brain Tumor Segmentation Challenge (BraTS) datasets, we demonstrate the efficacy of our SR method in application of spatial resolution enhancement of brain tumor MRI images to potentially maintain crucial diagnostic information for further clinical tasks.



\section{METHODS}
\label{sec:methods}


\subsection{Lesion Focused SR}
\label{ssec:roi}

\begin{figure} [!htbp]
   \begin{center}
   \begin{tabular}{c} 
   \includegraphics[width=\textwidth]{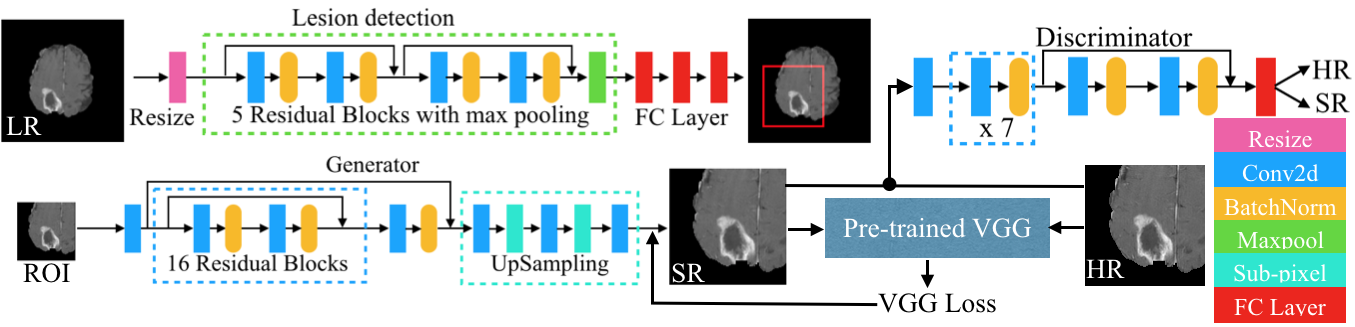}
   \end{tabular}
   \end{center}
   \caption[ldnet] 
   { \label{fig:ldnet} 
The schema of our proposed lesion focused super resolution (LFSR) neural network.
}
   \end{figure}

Our LFSR includes a lesion detection neural network $\mathrm{LD}$, a super resolution images generator $G$, a HR/SR images discriminator $D$, and a pre-trained 19 layers VGG\cite{vgg19} (Fig.\ref{fig:ldnet}). The $\mathrm{LD}$ aims to detect the region of interest (ROI, e.g. brain tumors), $I_\mathrm{lr}$ and $I_\mathrm{hr}$, from whole size LR and HR images $I_\mathrm{LR}$ and $I_\mathrm{HR}$ before we applying the GAN:
\begin{equation}
I_\mathrm{lr, hr} = \mathrm{LD}(I_\mathrm{LR, HR})
\end{equation}
We propose a max pooling residual block and an input-scale free residual neural network $\mathrm{LD}$. Compared to the residual blocks \cite{resnet} and skip connection have been widely used, a max pooling layers is added after two residual blocks, which include two skip connections between four convolution and batch normalization layers. This can help accelerate the training process, and reduce the memory cost of the ROI detection task.

During training, $G$ and $D$ of the GAN are playing a game: $G$ aims to estimate as realistic as possible SR images, $I_\mathrm{SR}$, from $I_\mathrm{LR}$, and the discriminator aims to figure them out from the ground truth $I_\mathrm{HR}$. With the lesion detection, the training aims to solve:
\begin{equation}
(\hat{\theta_\mathrm{G}}, \hat{\theta_\mathrm{D}}) = \operatorname*{argmin}_{\theta_\mathrm{G}, \theta_\mathrm{D}} \sum l^\mathrm{G}(G_{\theta_\mathrm{G}}(I_\mathrm{lr}), I_\mathrm{hr}) + l^\mathrm{D}(D_{\theta_\mathrm{D}}(G_{\theta_\mathrm{G}}(I_\mathrm{lr}), I_\mathrm{hr})))
\end{equation}
where $\hat{\theta_\mathrm{G}}$ and $\hat{\theta_\mathrm{D}}$ are the trainable parameters, $l^\mathrm{G}$ and $l^\mathrm{D}$ are the loss functions for the $G$ and $D$. In our proposed LFSR, we use a SR residual network (SRResNet) as the generator $G$ , which includes 16 residual blocks, and following sub-pixel convolution layers. The discriminator $D$ and the pre-trained VGG are trained simultaneously with $G$ to generate perceptually realistic image features.

\subsection{Data Preprocessing and Training Settings}
\label{ssec:data}
We have tested bilinear interpolation, SRResNet, SRGAN\cite{ledig2017photo} and LFSR on the post-contrast T1-weighted (T1Gd) MRI scans from the BraTS 2018 datasets \cite{menze2015multimodal}, which have been randomly divided into training ($9559$ images) and validation ($2368$ images) datasets. All the slices are normalized to zero-mean and unit-variance. We simulated the LR images by downsampling the HR ground truth and tested with additive white Gaussian noises (AWGN, $\sigma=20, 40$) applied in the \emph{k}-spaces \cite{Bao2003}. 

All the experiments were performed on a Linux workstation with NVIDIA TITAN Xp GPUs. All the models were implemented in Python, based on the TensorLayer \cite{tensorlayer2017} library, and were trained with Adam optimizer with the initial learning rate of $10^{-4}$. The $\mathrm{LD}$ was trained independently for 100 epochs with $L_\mathrm{2}$ loss. The SRResNet was trained for 350 epochs with the pixel-wise mean square error loss $l_\mathrm{MSE}$. The generator in SRGAN and LFSR was initially trained with $l_\mathrm{MSE}$ for 50 epochs, then the GAN was trained with $l^\mathrm{G} = l_\mathrm{MSE} + l_\mathrm{VGG} - l_\mathrm{GD}$, and $l^\mathrm{D}=1-l_\mathrm{DT}-l_\mathrm{DF}$, where $l_\mathrm{GD}$ was the percentage of incorrectly distinguished $I_\mathrm{{SR, sr}}$, and $l_\mathrm{{DT}}$, and $l_\mathrm{{DF}}$ were the percentages of correctly distinguished $I_{\mathrm{HR,hr}}$ and $I_{\mathrm{SR, sr}}$.

\section{RESULTS AND DISCUSSIONS}
\label{sec:results}

\subsection{Lesion Detection}
\label{ssec:roirst}
Our $\mathrm{LD}$ has achieved high accuracies on both X2 and X4 downsampled images. In evaluation, we defined that if a tumor was $100\%$ covered by the predicted ROI, it was a perfect detection, and if it was $95\%$ covered, it was a acceptable detection. In the X2 case, $2218$ images ($93.7\%$) were perfect detections, and other $111$ ($98.4\%$) were acceptable detections. In the X4 case, $2109$ images ($89.1\%$) were perfect detections, and other $119$ ($94.1\%$) acceptable detections . 

\subsection{X2 and X4 SR}

\begin{figure} [!htbp]
   \begin{center}
   \begin{tabular}{c} 
   \includegraphics[width=\textwidth]{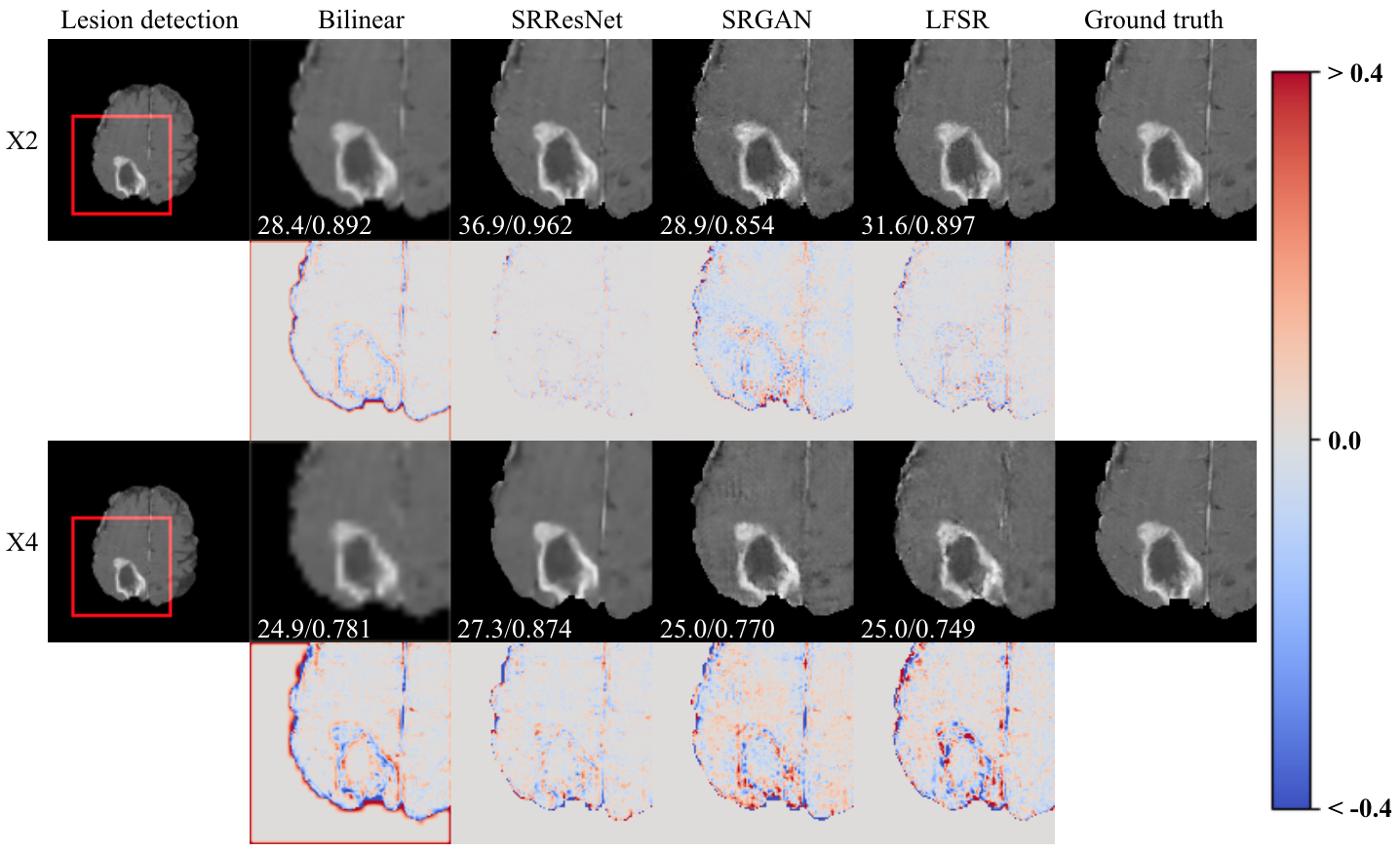}
   \end{tabular}
   \end{center}
   \caption[x2x4rst] 
   { \label{fig:x2x4roi} 
The ROIs of ground truth, the detected ROI with the predicted SR images(range: [-1, 1]), PSNR/SSIM of each result are also displayed.}
   \end{figure}

Here we showed the X2 and X4 SR  results in Fig. \ref{fig:x2x4roi}. Both bilinear interpolation and SRResNet have produced blurry SR results although SRResNet has achieved the highest PSNR. SRGAN and our proposed LFSR have resulted in images with more realistic texture features compared to the ground truth. Compared to the SRGAN, our LFSR has obtained higher (X2 cases) or equivalent (X4 cases) PSNR. More importantly, our LFSR has achieved significant reduction of the GPU memory cost; therefore, our LFSR can double the batch size, which has accelerated the training process to 266.8s/epoch for X2 and 194.8s/epoch for X4 (compared to SRGAN training time 649.8s/epoch for X2, and 370.8s/epoch for X4).   

\subsection{X2 SR with Additive Noise}
\label{ssec:denoise}
We have also tested LFSR in X2 SR with additive Gaussian noise. The bilinear interpolation with non local means denoising \cite{nonlocaldenoising} method (B+NLD) was tested to suppress the noise and provided a more fair comparison. All three deep learning methods have achieved higher PSNR and SSIM when noise presented (Table \ref{tabel:rst}). The $l_\mathrm{MSE}$ based SRResNet has still achieved the highest PSNR and SSIM. In contrast, both SRGAN and LFSR have been still able to generate more perceptually realistic textures from our qualitative studies. Furthermore, LFSR has achieved higher PSNR and SSIM than the SRGAN for the noisy cases (Table \ref{tabel:rst}), and more efficient training.

\begin{table}[!htbp]
\centering
\caption{PSNR and SSIM results for simulations with and without additive Gaussian noise (bold: better than SRGAN).}
 \begin{adjustbox}{width=\textwidth}

\begin{tabular}{cl|l|l|l|l|l|l|l}
\hline
\multicolumn{1}{l}{\multirow{2}{*}{}} & \multicolumn{4}{c|}{PSNR}                                                                              & \multicolumn{4}{c}{SSIM}                                                                        \\ \cline{2-9} 
\multicolumn{1}{l}{}                  & \multicolumn{1}{c|}{X2} & \multicolumn{1}{c|}{X4} & \multicolumn{1}{c|}{X2($\sigma = 40$)} & \multicolumn{1}{c|}{X2($\sigma = 40$)} & \multicolumn{1}{c|}{X2} & \multicolumn{1}{c|}{X4} & \multicolumn{1}{c|}{X2($\sigma = 20$)} & \multicolumn{1}{c}{X4($\sigma = 40$)} \\ \hline
\multicolumn{1}{c|}{B+NLD}            & 29.1              & 25.2              & 20.7                    & 17.1                    & 0.900             & 0.761             & 0.623                   & 0.483                  \\
\multicolumn{1}{c|}{SRResNet}         & 35.6              & 27.9              & 32.8                    & 30.6                    & 0.962             & 0.832             & 0.895                   & 0.840                  \\ \hline
\multicolumn{1}{c|}{SRGAN}            & 29.6              & 25.7              & 27.6                    & 26.2                    & 0.865             & 0.741             & 0.789                   & 0.731                  \\
\multicolumn{1}{c|}{LFSR}             & \textbf{32.1}     & 25.1              & \textbf{29.0}           & \textbf{27.4}           & \textbf{0.914}    & 0.723             & \textbf{0.832}          & \textbf{0.772}        
\end{tabular}
\label{tabel:rst}
\end{adjustbox}
\end{table}

\section{CONCLUSION}
\label{sec:conclusion}

In summary, we have developed and validated a lesion focused SR (i.e., LFSR) method to super-resolve the tumor ROIs imaged by MRI. Compared to state-of-the-arts SR method, our proposed LFSR method is more efficient and it can result in perceptually more realistic SR, which will maintain crucial image features for further clinical tasks and decisions. In the final camera ready version, we will include a more detailed description of our method and more comparison results.

\bibliography{report} 
\bibliographystyle{spiebib} 

\end{document}